\documentclass[conference,10pt]{IEEEtran}
\IEEEoverridecommandlockouts

\usepackage{cite}
\usepackage{amsmath,amssymb,amsfonts}
\usepackage{graphicx}
\usepackage{tikz}
\usepackage{pgfplots}
\usepackage{pgfplotstable}
\usepackage{ifthen}
\pgfplotsset{compat=1.16,set layers, mark layer=axis tick labels}
\usepackage{soul}
\usepackage[font=footnotesize]{subcaption}
\usepackage{tabularx}
\usepackage[font=footnotesize]{caption}
\usepackage{textcomp}
\def\BibTeX{{\rm B\kern-.05em{\sc i\kern-.025em b}\kern-.08em
    T\kern-.1667em\lower.7ex\hbox{E}\kern-.125emX}}
\usepackage{amsmath}
\usepackage{amssymb}
\usepackage{optidef}
\usepackage{algorithm}
\usepackage{enumitem}
\usepackage{hyperref}
\usepackage{xcolor} 
\usepackage{mathtools}
\usetikzlibrary{plotmarks}

\usepackage{algpseudocode}
\newcommand*\Ex[1]{\mathbb{E}\begin{Bmatrix} #1 \end{Bmatrix}}

\newcommand*\mb[1]{\mathbf{#1}}
\newcommand*\bs[1]{\boldsymbol{#1}}

\pgfplotsset{every axis/.append style={
		scaled x ticks = false,
		label style={font=\footnotesize},
		tick label style={font=\footnotesize},
		tick scale binop=\times
	}
}
\pgfkeys{/pgf/number format/.cd,%additional code
	1000 sep={},%additional code
}

\title{Cell-Free Massive MIMO under a Non-Linear Power Amplifier Consumption Model}
\author{\IEEEauthorblockN {Robbert Beerten, Vida Ranjbar, Hazem Sallouha, Sofie Pollin \vspace{-0.8cm}}
\thanks{R. Beerten, V. Ranjbar, H. Sallouha and S. Pollin are with the Department of Electrical Engineering of KU Leuven, Belgium (corresponding author: \{robbert.beerten\}@kuleuven.be). This research has received funding from the Smart Networks and Services Joint Undertaking (SNS JU) under the European Union’s Horizon
Europe research and innovation programme under Grant Agreement No. 101096954 (6G-BRICKS) and 101139257 (6G-SUNRISE) and
from Research Foundation – Flanders (FWO) project number G0C0623N. 
}}

\begin{document}
\maketitle%
\begin{abstract}
    Existing works on Cell-Free Massive MIMO primarily focus on optimising system throughput and energy efficiency under high-traffic scenarios with only a limited focus on variable user demand as required by higher network layers. 
   Additionally, existing works only minimise the transmitted power instead of the consumed power at the power amplifier.
   This work introduces a penalty-method-based approach to minimise the amplifier's power consumption while scaling much better with network size than current solutions and promoting sparsity in the power allocated to each access point. Furthermore, we demonstrate substantial reductions in power consumption (up to 24\%) 
   by considering the non-linear power consumption. 
\end{abstract}
\section{Introduction}
 Cell-Free Massive MIMO (CF mMIMO) considers a network architecture where users are served by many densely deployed Access Points (APs) via coherent transmission and reception \cite{Demir_Bjornson_Sanguinetti_2021}. This dense deployment of APs ensures small AP-user distances and significantly reduced path losses between users and their nearest APs. CF mMIMO promises a more energy-efficient operation than classic cellular architectures in high-traffic scenarios by optimally allocating transmission power to each AP-user connection. However, to achieve this optimal energy efficiency, a Central Processing Unit (CPU) must solve a high-dimensional transmit power allocation problem with channel knowledge between all APs and users. These complex problems create major scalability challenges in practical scenarios and render many state-of-the-art solutions infeasible for practical implementation \cite{Mai_Ngo_Tran_2022}.

 Existing literature has focused on optimizing throughput and energy efficiency while assuming a high traffic demand. However, it is well-known that traffic load in broadband networks is highly variable \cite{Auer_Giannini_Desset_Godor_Skillermark_Olsson_Imran_Sabella_Gonzalez_Blume_etl} and in low-demand periods it is crucial to minimize power consumption while maintaining per-user Quality-of-Service (QoS) requirements \cite{Piovesan_Lopez-Perez_De_Domenico_Geng_Bao_Debbah_2022}.
 %\cite{Piovesan_Lopez-Perez_De_Domenico_Geng_Bao_Debbah_2022}
To address this issue, recent works proposed methods that focus on several variations of downlink resource allocation problems with per-user Signal-to-Interference-and-Noise-Ratio (SINR) requirements
\cite{Demir_Masoudi_Bjornson_Cavdar_2024, eucncrobbert, Persson_Eriksson_Larsson_2013, Rottenberg_2024,Peschiera_Rottenberg_2023, Farooq_Ngo_Hong_Tran_2021}.
However, these existing works still suffer from two main shortcomings: 

\textbf{P1:}
Most current research uses simplified power amplifier (PA) models that assume power consumption increases linearly with transmission power. In reality, this relationship is non-linear, 
% \cite{Mikami_Takeuchi_Kawaguchi_Ohta_Yoshimoto_2007}
leading to suboptimal performance in real-world settings.
Despite this issue, only very few works have addressed this shortcoming. For example, Persson \textit{et al.} \cite{Persson_Eriksson_Larsson_2013} focus on power minimization over a general MIMO channel, while Rottenberg \cite{Rottenberg_2024} addresses energy-efficient transmission over SISO channels under different realistic PA power models. Peschiera \textit{et al.} \cite{Peschiera_Rottenberg_2023} examine power minimization for a centralized Massive MIMO base station. 
% Our work is the first to address realistic non-linear PA consumption models in CF mMIMO networks. We significantly reduce the network's overall power consumption if the model is accurately accounted for.

\textbf{P2:} 
% Existing works consider joint precoding and power allocation and often rely on zero-forcing constraints, essentially removing inter-user interference. However, in the CF mMIMO networks, precoding is usually computed at the AP and power allocation at the CPU. As a result, the users still suffer from interference,Thesescenario; thus 
Downlink transmit power allocation in CF mMIMO leads to a high-dimensional and highly coupled problem, where allocating more power at one AP to a single user increases interference for all other users. 
When each user must attain a certain SINR, this coupling becomes highly important 
% since the ratio between signal power and interference plus noise becomes essential 
and requires solving a Second-Order Cone Programming (SOCP) problem.  
% Many works only formulate these SOCP problems and then solve them via conventional commercial convex solvers. 
The scalability issue of solving such SOCP problems in large networks is a major drawback. To this end, recent works \cite{Farooq_Ngo_Hong_Tran_2021, Mai_Ngo_Tran_2022} have explored first-order optimization methods for CF mMIMO % \vida{You mentioned CFmMIMO in the beginning, but defined it here, maybe better to define it in the beginning where you first used it?}
networks. However, these works did not address the non-linearity in the PA's power consumption. 
% In this work, we overcome the limitations of SOCP-based solutions by introducing a low-complexity, penalty-method-based approach that efficiently handles large-scale CF mMIMO networks while handling the non-linearity of the PAs' power consumption.

Considering the two aforementioned fundamental problems, our work's contribution is twofold, as detailed in the following: % \vida{I forgot what C1 and C2 stand for, I see that they are describing the solutions provided for P1 and P2, but why you chose C1 and C2 naming? also well-written introduction}

\textbf{C1:} %We formulate the power minimisation problem based on local per-AP distributed precoding along with a realistic power consumption model of the PA in CF mMIMO. 
We propose to account for the increasing energy efficiency of the PA, which increases with power output 
% \vida{output power you mean? Also, you can only say energy efficiency of PA without "with increased output power, I did not understand the second part."} 
when minimising the consumed power at the PA. We show that this approach saves significant power in the PA. Furthermore, the approach saves relatively more power when the network has a low traffic demand, a scenario that has been heavily understudied in the literature. Additionally, we show that optimizing the power allocation for the non-linear power model naturally induces sparsity in the per-AP allocated power, thus implying several APs must shut down for optimal PA power consumption, even without considering the fixed power consumption of the APs.

\textbf{C2:}  We propose a penalty-method-based approach that efficiently scales to large-scale networks. Our method significantly outperforms current generic SOCP solvers regarding computational scalability and allows for implementation in networks with a massive number of APs.

\textbf{Notation:} 
% A diagonal matrix with the elements $x_i$ on the main diagonal is denoted by $\text{diag}(x_0, x_1, \cdots, x_N)$. 
The $l$'th element of vector $\mathbf{a}$ is denoted by $[\mathbf{a}]_l$. 
% The expected value of $x$ is denoted by $\Ex{x}$. 
The complex multivariate normal distribution, with covariance $\mb{A}$ and zero mean is denoted by $\mathcal{CN}(\mb{0}, \mb{A})$.
% The notations $\mathbb{R}^{M \times N}, \mathbb{R}^{N}_{\succeq \mb{0}}$ and $\mathbb{C}^{M \times N}$ denote the set of real $M$ by $N$ matrices, real $N$ dimensional vectors with non-negative elements and complex $M$ by $N$ matrices, respectively. 
The basis vector of length $A$ with index $a$ as the non-zero element is indicated by $\mb{e}_{A}^a$. The Kronecker product is denoted by $\otimes$. The function $\max(0,x)$ is represented by $[x]_+$.
\section{System Model}

\label{sec:sysmodel}
We consider a system with $K$ single-antenna users, $L$ APs with $N$ antennas each, and a single CPU. 
We model the wireless channel between each user $k$ and AP $l$ as Rayleigh fading, which assumes a rich scattering environment. The channel $\mathbf{h}_{lk} \in \mathbb{C}^N$, is modelled as a realization of a multivariate random distribution, denoted as
$\mathbf{h}_{lk} \sim \mathcal{CN}(\mb{0}, \mb{R}_{lk})$. The covariance is generated via the one-ring scattering model \cite{bjornsonMassiveMIMONetworks2017}. 
We model the received signal at user $k$ as \cite{Demir_Bjornson_Sanguinetti_2021}: 
\begin{equation}
    y_k = \sum_{l=1}^L  \sum_{i=1}^{K} \rho_{li} \mb{h}_{lk}^H \mb{w}_{li} s_i + n_k,
    \end{equation}
where $\rho_{lk}$ and $\mb{w}_{l,k}$ are the DL power and DL precoder for user $k$ at AP $l$, respectively, $n_k$ is the thermal
noise at user $k$ and is generated from the distribution $\mathcal{CN}(0, \sigma_{\text{DL}}^2)$. 
The signal transmitted to user $k$, $s_k$, is assumed to be from the distribution  $\mathcal{CN}(0, 1)$ and mutually uncorrelated amongst different users. 
We consider a reciprocal channel in the UL and DL, and for its estimation, we assume each user has a distinct pilot sequence. 
Accordingly, each AP locally estimates the UL channel as \cite{Demir_Bjornson_Sanguinetti_2021},
\begin{equation}
        \hat{\mb{h}}_{lk} = \sqrt{\tau_p p_k} \mb{R}_{lk} (\tau_p p_k\mb{R}_{lk} + \sigma_{\text{UL}}^2\mb{I}_N)^{-1} \mathbf{y}^{(p)}_{lk}\,, 
\end{equation}
%\vida{So you assumed pilot power is 1 in (1)?}
where $\sigma_{\text{UL}}^2$ is the thermal noise power at AP $l$, $p_k$ is the power at which user $k$'s pilot is transmitted, $\tau_p$ is the length of the pilot and $\mathbf{y}^{(p)}_{lk}$ is the decorrelated pilot symbol at AP $l$ for user $k$. Each AP locally computes local partial MMSE precoders as \cite{Demir_Bjornson_Sanguinetti_2021}:
\begin{equation}
    \mb{w}_{l,k} = \left(\sum_{i \in \mathcal{S}_l} p_k (\hat{\mb{h}}_{l,i}\hat{\mb{h}}_{l,i}^H + \mb{C}_{l,i}) + \sigma_{\rm{DL}}^2 \mb{I}_N\right)^{-1} \hat{\mb{h}}_{l,k},
\end{equation}
% \vida{You are computing the precoding vector for user k, right? but in you summation maybe better use another index like $i\in \mathcal{S}_l$? also why do you say $\mathcal{S}_l$ is the nearest users, is not it the set of served users as well?}
where $\mathcal{S}_l$ refers to the users with the largest channel gain to AP $l$. 
Symbol $\mb{C}_{l,k}$ indicates the covariance matrix of the channel estimation error \cite{Demir_Bjornson_Sanguinetti_2021}.
% i.e., $\mb{C}_{l,k} = \Ex{\hat{(\mb{h}}_{l,k} - \mb{h}_{l,k}) (\hat{\mb{h}}_{l,k} - \mb{h}_{l,k})^H}$. 
The symbol $p_k$ refers to the power at which the uplink pilot was transmitted by user $k$.
% \begin{equation}
% \begin{aligned}
%     y_k &= \sum_{l=1}^L\rho_{lk}  \Ex{\mb{h}_{lk}^H \mb{w}_{lk}}  s_k  + \sum_{l=1}^L\sum^{K}_{i=1} \rho_{li} \mb{h}_{lk}^H \mb{w}_{li}  s_i \\
%     &  - \sum_{l=1}^L \rho_{lk}  \Ex{\mb{h}_{lk}^H \mb{w}_{lk}} s_k  + n_k.
%     \end{aligned}
% \end{equation}
Similarly to the analysis in \cite[Sec. 7.3]{Demir_Bjornson_Sanguinetti_2021} a practically achievable downlink SINR is then:  
% \begin{equation}
% \begin{aligned}
%     & \gamma_k =\\ 
%     &\frac{| \sum\limits_{l=1}^L\sqrt{\rho_{lk}} \Ex{\mb{h}_{lk}^H \mb{w}_{lk}} |^2}
%     {\sum\limits_{i=1}^K \Ex{|\sum\limits_{l=1}^L \sqrt{\rho_{li}} \mb{h}_{lk}^H \mb{w}_{li}|^2} - |\sum\limits_{l=1}^L \sqrt{\rho_{lk} }\Ex{\mb{h}_{lk}^H \mb{w}_{lk}}|^2 + \sigma_{DL}^2}\,, \\
%     \end{aligned}
% \end{equation}
%  which can be simplified to the following expression,
\begin{equation}
    \gamma_k = \frac{|\mb{b}_k^T\bs{\rho}_{k}|^2}{\sum\limits_{i=1}^K\bs{\rho}_i^{T}\mathbf{C}_{ki}\bs{\rho}_i - |\mb{b}_k ^{T}\bs{\rho}_{k}|^2 + \sigma_{DL}^2} \,,
    \label{eqn:ErgRate}
\end{equation}
where we introduce the vector $\bs{\rho}_k = [\rho_{1k} \; \rho_{2k} \; \dots \; \rho_{Lk}]^T$ that contains the DL power coefficients between the $k$-th user and each AP. Moreover, $\mb{b}_k  \in \mathbb{R}_{\succeq \mb{0}}^{L} $
and $\mb{C}_{ki}  \in \mathbb{C}^{L \times L}$ denote, 
\begin{subequations}
    \begin{align}
        [\mathbf{b}_k]_l &= \Ex{ \mb{h}_{lk}^H \mb{w}_{lk}} \,, \\ 
        [ \mathbf{C}_{ki}]_{l,m} &= \Ex{\mb{h}_{lk}^{H}\mb{w}_{li}\mb{w}_{mi}^{H} \mb{h}_{mk}},
    \end{align}
\end{subequations}
for notational brevity and ease of exposition similarly to \cite{Demir_Masoudi_Bjornson_Cavdar_2024}. 
Furthermore $\mb{C}_k$ combines all $\{ \mb{C}_{ki} \}_i$ in a blockdiagonal matrix.
In the rest of this paper, when using the term Spectral Efficiency (SE), we refer to the achievable downlink SE, which is $\mathrm{SE}_k = \log_2(1+\gamma_k)$.
This relation is important as we will use both SINR and SE throughout this paper.

\subsection{Power Consumption Model}
In this subsection, we develop our power consumption model.
As mentioned in \textbf{P1}, most state-of-the-art works rely on heavily idealised PA models where the consumed and transmitted power have a linear relation. However, it has been shown that this relation is, in fact, nonlinear \cite{Persson_Eriksson_Larsson_2013}. In this section, we elaborate on a more accurate power consumption model.
% \vida{you did not define $diag(\mathbf{e}_L^l)$ here.}
The total transmitted power at AP $l$ is written as, 
\begin{equation}
     P^{\text{tx}}_l = \sum_{k=1}^{K} \rho_{lk}^2.
\end{equation}
Using this notation, the consumed power can be calculated as a function of the transmitted power via one of two models: an \textbf{Ideal} % \vida{So ideal mean linear mentioned in previous paragraph?}
%\vida{before you used idealized, maybe better use either ideal or idealized every where,}
PA model or a \textbf{Non-Linear} PA model. The model for the ideal PA assumes a linear relation between the transmitted and consumed power $P^{\text{ideal}}$ with a constant efficiency $\eta$. On the other hand, the non-linear model considers a variable efficiency $\eta \leq \eta_{max}$, which increases with transmitted power. % \vida{it is not evident from the formula 8 that $\eta_{max}$ increases with transmitted power right?, because when transmitted power increases, consumed power also increase, so I don't see immediately how $\eta_{max}$ increases. If there is a ref for that, maybe better cite it here.}:
\begin{equation}
   \eta = \frac{P^{\text{tx}}_l }{P^{\text{non-linear}}_l} = \eta_{\text{max}} \left( \frac{ P^{\text{tx}}_l }{P_{\text{max}}} \right)^{\frac{1}{2}}.
\end{equation}
It has been shown that this model is quite accurate when working sufficiently far from the saturation point of the class B PA. % \cite{He_Srikanteswara_Bae_Newman_Reed_Tranter_Sajadieh_Verhelst_2011}. 
We denote this point, sufficiently far away from the saturation point, by $P_{\text{max}}$. The model was also considered in \cite{Rottenberg_2024, Peschiera_Rottenberg_2023, Persson_Eriksson_Larsson_2013}. 
Finally, we consider two per-AP power consumption models, an ideal power consumption model, $P^{\text{ideal}}_l(P^{\text{tx}}_l)$, and a non-linear power consumption model, $P^{\text{non-linear}}_l(P^{\text{tx}}_l)$:
\begin{align}
     \textbf{Ideal PA: \;} & P^{\text{ideal}}_l(P^{\text{tx}}_l) = \frac{1}{\eta}   P^{\text{tx}}_l \label{eq:paIdeal} \\
     \textbf{Non-Linear PA: \;} & P^{\text{non-linear}}_l(P^{\text{tx}}_l) =  \frac{1}{\eta_{\text{max}}}  \sqrt{P^{\text{tx}}_l P_{\text{max}}}.
     \label{eq:paReal}
     \end{align}
    %  \vida{It took me time to see $P_l^{ideal}$ is the power consumption under linear model, maybe you can use $P_l^{cons-L}$ and $P_l^{cons-NL}$ for linear and non-linear power consumption notation, or something that is less time consuming to guess?}
% This second expression suggests that a PA becomes more efficient as the transmitted power approaches $P_{\text{max}}$.
% \vida{Or, at maximum transmitted power, the realistic model is same as ideal PA model. also $\eta$ should be $<1$, meaning that the consumed power is always larger than the transmit power} 
The non-linear model more accurately reflects real-world PA behaviour and enables us to save significant energy by directly minimising the consumed power instead of just minimising the transmitted power. We will show later that interestingly, due to the inherent far-near effect of users in CF mMIMO networks and the relatively higher penalty on small transmit powers in the non-linear model, the final result only allocates a non-zero transmit power to a limited subset of APs in the network when the demand is relatively low.  
 
\subsection{Downlink Power Allocation}
 Regarding downlink power allocation, we consider the following problem: minimising the consumed power at the APs' PAs while serving users under an SINR constraint for each user and per AP power constraints: 
% \begin{equation}
    \begin{subequations}
    \begin{align}
        \mathcal{P}1:  {\min_{\rho_{l,k} \geq 0}}  & \quad \sum_{l=1}^{L} P^{\text{non-linear}}_l( P_l^{\text{tx}}) \\
         \text{s.t.} \qquad \gamma_k &\leq \frac{|\mb{b}_k^T\bs{\rho}_{k}|^2}{\sum\limits_{i=1}^K\bs{\rho}_i^{T}\mathbf{C}_{ki}\bs{\rho}_i - |\mb{b}_k ^{T}\bs{\rho}_{k}|^2 + \sigma_{DL}^2}, & \forall k \label{con:noSlack} \\ 
          \sqrt{P_{\text{max}}} 
        &\geq \| \rho_{l1} \; \dots \; \rho_{lK} \|_2 , & \forall l  {\label{con:mixer}}.
        \end{align}
    \end{subequations}
% \end{equation}
% \begin{mini!}[3]
%  	{\rho_{l,k} \geq 0}
%     {\sum_{l=1}^{L} P^{\text{non-linear}}_l( P_l^{\text{tx}})}
%     {}
%     {\mathcal{P}1:}  
%     \addConstraint
%     { \gamma_k }
%     {\geq \frac{|\mb{b}_k^T\bs{\rho}_{k}|^2}{\sum\limits_{i=1}^K\bs{\rho}_i^{T}\mathbf{C}_{ki}\bs{\rho}_i - |\mb{b}_k ^{T}\bs{\rho}_{k}|^2 + \sigma_{DL}^2} \qquad }
%     {\forall k}
%         \addConstraint
%    }
%     {\forall l}\,.
% \end{mini!}

% By having formalized the exact problem, we can now make our contribution more explicit: 

% \begin{enumerate}
% \item[\textbf{C1:}] We directly minimise the consumed power by incorporating our realistic power consumption model in the cost function of $\mathcal{P}1$.  
% \item[\textbf{C2:}] We develop a low-complexity approach for finding highly accurate approximate solutions of $\mathcal{P}1$. 
% \end{enumerate}
We introduce extra notations to simplify the mathematical exposition. First, let $\mb{x}$ denote the stacked version of the power control coefficients, i.e. $\mb{x} = [\bs{\rho}_{1}^T \; \bs{\rho}_{2}^T \; \dots \; \bs{\rho}_{K}^T ]^T$. Second, we isolate the power coefficients of AP $l$ from that variable $\mb{x}$ as,
\begin{equation}
    \mb{x}_l =  ( \mb{I}_K  \otimes \text{diag}(\mb{e}_L^l) ) \;\mb{x} \,.
\end{equation}
Thirdly, we align  $\mb{b}_k$ with ${\mb{x}}$ as,
\begin{equation}
    \tilde{\mb{b}}_k = \mb{e}^k_K \otimes  \mb{b}_k. 
\end{equation}
Finally, we reformulate the constraint (\ref{con:noSlack}) as in \cite{eucncrobbert}:
% \vida{shouldnot $\mathbf{b}^T_k\mathbf{\rho}_k$ be inside abs? $||$} \robbert{$\mb{b}$ and $\bs{\rho}$ are both non negative} 
\begin{equation}
\begin{aligned}
 % & \frac{|\mb{b}_k^T\bs{\rho}_{k}|^2}{\sum\limits_{i=1}^K\bs{\rho}_i^{T}\mathbf{C}_{ki}\bs{\rho}_i - |\mb{b}_k ^{T}\bs{\rho}_{k}|^2 + \sigma_{DL}^2} \geq  \gamma_k\\ 
\sqrt{\frac{1 + \bar{\gamma}_k }{\bar{\gamma}_k }}\mb{b}_k ^{T}\bs{\rho}_{k}
    & \geq
    \| [ (\mathbf{C}_{k1}^{\frac{1}{2}}\bs{\rho}_1)^T \: \dots \:  (\mathbf{C}_{kK}^{\frac{1}{2}}\bs{\rho}_K)^{T}  \; \sigma_{DL}]^T \|_2  \\ 
    0
     & \geq 
      \| (\mb{C}_k^{\frac{1}{2}} \mb{x})^T  \; \sigma_{DL} \| -   \sqrt{\frac{1 + \gamma_k }{\gamma_k}} \tilde{\mb{b}}_k^T \mb{x}  \\ 
       0  &\geq g_k(\mb {x}),
     \end{aligned}
\end{equation}
% \vida{where is the 11c constraint?}
where we introduce $ g_k(\mb {x})$ as the constraint violation.
 
\section{Proposed PA Power Minimisation}
% This section explains how to tackle \textbf{P2}. 
Generally, convex formulations such as $\mathcal{P}1$ are not solved explicitly in state-of-the-art works but via generic convex solvers such as CVX \cite{cvx}. However, as pointed out in \textbf{P2}, generic solvers suffer from poor scaling with increasing network size. 
Hence, we propose a low-complexity method for solving $\mathcal{P}1$ by penalty-based method in Algorithm~\ref{alg:penalty}. We subsequently provide an accelerated gradient method in Algorithm~\ref{alg:APG}  for solving the subproblems of the penalty method.

\subsection{Penalty Method} 
% This proves to be extremely useful as the SINR constraints usually require solving the problem by interior point methods, which scale poorly with network size (cf. \textbf{P2}) due to the required calculation of the Hessian matrix of the cost and constraint functions. 
The penalty method allows for finding approximate solutions to a constrained problem, such as $\mathcal{P}1$, by eliminating some constraints and incorporating them into the cost function via the following penalty function: 
\begin{equation}
        \Psi_k(\mb{x}) =  [g_k(\mb{x})]_+^2. 
\end{equation}
The penalty function ensures that any violations of the QoS constraints contribute to the total cost, thereby driving the solution toward feasibility if the penalty weight % \vida{you mean penalty weight?}
is sufficiently large. 
In fact, if the relative weight of the penalty goes to infinity, the solution of the penalised problem is equivalent to the constrained problem $\mathcal{P}1$. Unfortunately, directly solving the problem for very large penalty weights leads to an ill-conditioned problem. To alleviate this problem, we start from small values for the penalty weights and increase these iteratively until the QoS requirements are satisfied, whereby the solution of the previous iteration is the starting point for the next iteration.
Let $\lambda(i)$ be the penalty weight at iteration $i$ of the penalty method. 
The cost function of the penalty method for a penalty $\lambda(i)$, is then defined as follows: 
% \vida{where is the objective function of P1, summation of the consumed power in 14?} 
\begin{equation}
    f^{\lambda(i)}(\mb{x}) = \sum_{l=1}^{L} P^{\text{non-linear}}_l (\mb{x}) + \lambda(i) \sum_{k=1}^K  \Psi_k(\mb{x}) \,.
\end{equation}
The only remaining constraints are then (10c):
% \vida{In P2, Should we minimize also for $\lambda_k$? also I don't understand the objective of P2, it should have sum of consumed power as in P1, no? Also, why do we have two algorithms? Is Alg2 a benchmark? as far as I remember, you wanted to prove that algorithm 1 converges?}
% \begin{mini!}[3]
%   {\mb{x} \geq 0}
%     {f^{\bs{\lambda}(i)}(\mb{x})} 
%     {}
%     {\mathcal{P}2:}{ }  
%     \addConstraint
%     {\sqrt{P_{max}}}
%     {\geq \|   \mb{x}_l  \|_2 }
%     {\; \forall l}\,.
% \end{mini!}
    \begin{subequations}
    \begin{align}
        \mathcal{P}2:  {\min_{\mb{x} \geq 0}}  & \quad f^{\lambda(i)}(\mb{x}) \\
         \text{s.t.} \qquad \sqrt{P_{\text{max}}} & \geq \|   \mb{x}_l  \|_2  & \forall l 
        \end{align}
    \end{subequations}
Algorithm \ref{alg:penalty} outlines how $\mathcal{P}2$ is solved sequentially to produce an approximate solution to $\mathcal{P}1$. The reader is referred to \cite{bentalOpt2lectures} for a detailed convergence proof of the penalty method.
 
\begin{algorithm}[H]
    \caption{Penalty Method}
    \begin{algorithmic}[H]
    \State{Initialize uniform random $\mb{x}^{0}$ in $[0, 10^{-10}]$, $\lambda(0) = 0.1$}
    \For{$i = 1 \; \dots I \; $}
        \State{$\mb{x}^{i} \gets \arg \min \mathcal{P}2$ with $\mb{x}^{i-1}$ as initial estimate}
        \If{$\Psi_k(\mb{x}^{i}) \simeq 0 \; \forall k$}
            \State{Terminate Algorithm}
        \EndIf
        \State{$\lambda(i+1) \gets \lambda(i) \zeta $}
        \EndFor
    \end{algorithmic}
    \label{alg:penalty}
\end{algorithm}

Note that the initialization point for $\mb{x}^{0}$ is chosen empirically here, future work might want 
to investigate this starting point. 

% \newcounter{mytempeqncnt}
% \begin{figure*}[htb]
%     \begin{equation}
%             \nabla f (\mb{x}) = \sum_{l=1}^L \frac{\sqrt{ P_{\text{max}}}}{ 2 \eta_{\text{max}} \sqrt{P_{\text{max}}  \| \mb{x}_l \|^2}} \mb{x}_l  
%             + 
%             \sum_{k=1}^K\lambda_k \max(0, \Psi_k(\mb{x})) 
%             \left( 
%             \frac{(\mb{C}_k + \mb{C}_k^T)}{2 \sqrt{\mb{x}^T \mb{C}_k \mb{x} + \sigma_{DL}^2}} 
%            \mb{x}- \sqrt{\frac{1 + \gamma_k }{\gamma_k }}\tilde{\mb{b}}_k 
%             \right)
%             \label{eq:gradient}
%     \end{equation}
%     \hrulefill
% \end{figure*}
\vspace{-0.15cm}
\subsection{Gradient Descent}
If $\mathcal{P}2$, is smooth and strongly convex it can be solved by the Accelerated Projected Gradient (APG) method \cite{Mai_Ngo_Tran_2022}. 
The proof for smoothness of the penalties $\Psi_k(\mb{x})$ can be inferred from a similar proof provided in \cite{Mai_Ngo_Tran_2022} but is omitted here due to space limitations. Unfortunately, the gradient of the accurate power model $\nabla P^{\text{non-linear}}_l (\mb{x})$, does not exist at $\| \mb{x}_l \|_2 = 0$.
\begin{equation}
 \nabla P^{\text{non-linear}}_l (\mb{x}) = \frac{\sqrt{ P_{\text{max}}}}{  \eta_{\text{max}}   \| \mb{x}_l \|_2} \mb{x}_l .
\end{equation}
% \vida{I think in equation (17), you should have $\sqrt{P_{max}}$ only in the numerator as for function $f=a\sqrt{x}$, $\frac{df}{dx}=\frac{a}{2\sqrt{x}}$}
This leads to non-smoothness if the power at one AP is pushed towards zero. To combat such effects, we use the smoothing approximation proposed by Nesterov \cite{Nesterov_2005} in the following section.
Furthermore, the APG method is not a pure descent method, even for convex minimisation. Since we rely heavily on aggressive early exiting from the gradient descent to minimize runtime, we prefer to use the monotone descent by Beck \textit{et al.} \cite{Beck_Teboulle_2009}. Algorithm~\ref{alg:APG} summarises this in pseudocode. 
\begin{algorithm}[htb]
    \caption{ Accelerated Gradient Descent}
    \begin{algorithmic}[htb]
    \State{\textbf{Input:} $\mb{x}^0,  \bs{\lambda}(i)$}
    \State{\textbf{Initialize:} $\mb{x}^1 = \mb{z}^1 = \mb{x}^0, t^1 = 1,$}
    \For{$t = 1$ \dots maxIterations}
        \State{$\mb{y}^{t} \gets \mb{x}^{t} + \frac{\mu^{t-1}}{\mu^{t}}(\mb{z}^{t} - \mb{x}^{t}) + \frac{\mu^{t-1}-1}{\mu^{t}}(\mb{x}^{t} - \mb{x}^{t-1})$}
        \State{$\mb{z}^{t+1} \gets \mathcal{P}_\mathcal{C}(\mb{y}^{t} - \alpha_t \nabla f(\mb{y}^{t}))$}
        \State{$\mu^{t+1} \gets \frac{1}{2}\sqrt{4 (\mu^{t})^2 + 1}+\frac{1}{2}$}
        \State{$\mb{x}^{t+1} \gets \begin{cases} \mb{z}^{t+1} & \text{if} \;f^{\bs{\lambda}(i)}(\mb{z}^{t+1}) \leq f^{\bs{\lambda}(i)}(\mb{x}^{t}) \\  
        \mb{x}^{(t)} & \text{else}
        \end{cases}$}
        \If{$ f^{\lambda^{(t)}}(\mb{x}^{t}) -  f^{\lambda^{(t)}}(\mb{x}^{t+1})  < \epsilon f^{\lambda^{(t)}}(\mb{x}^{t+1}) $}
            \State{Terminate Algorithm}
        \EndIf
    \EndFor
    \end{algorithmic}
     \label{alg:APG}
\end{algorithm}
% \vida{In alg2, I assume that $x^1$ should be from the feasible region or randomly generated? I think if you take all elements of $x^1$ equally such that the max power const is satisfied, then $x^1$ is indeed in the feasible region of max power. So with this assumption, I don't understand the step of $y^k$ and $t^k$ in alg2?}
% \robbert{The initialization is taken as input from alg 1. $y_k $and $t_k$ are actually for the acceleration of the gradient method. A good reference is this one \cite{Beck_Teboulle_2009}}
\subsection{Gradient Smoothing}
 According to the analysis in \cite{Nesterov_2005}, we smoothen the power consumption models around $ \| \mb{x}_l \|_2 \rightarrow 0$ as follows, 
\begin{equation}
    \psi_{\mu}(\| \mb{x}_l \|_2 ) = \begin{cases}
        0  & \| \mb{x}_l \|_2  = 0 \\ 
        \| \mb{x}_l \|_2 ^2 / 2\mu & 0 \leq \| \mb{x}_l \|_2  \leq \mu \\ 
        \| \mb{x}_l \|_2  - \mu/2 & \mu < \| \mb{x}_l \|_2  .
    \end{cases}
\end{equation}
This then leads to respective gradients,
\begin{equation}
    \nabla \psi_{\mu}(\| \mb{x}_l \|_2 ) =  
    \begin{cases}
     0  & \| \mb{x}_l \|_2  = 0 \\ 
           \mb{x}_l / \mu & 0 \leq \| \mb{x}_l \|_2  \leq \mu \\ 
          \mb{x}_l / \| \mb{x}_l \|_2  &   \mu < \| \mb{x}_l \|_2.
    \end{cases}
\end{equation}
We choose $\mu$ as $10^{-7}$ \footnote{The selection of $\mu$ also determines the error induced by the smoothing approximation. This error is upper bounded by $\mu/2$ \cite{Nesterov_2005}.}.
Finally the gradient of the smoothed power consumption model $\hat{P}^{\text{non-linear}}_l (\mb{x})$ is then, 
\begin{equation}
    \nabla \hat{P}^{\text{non-linear}}_l (\mb{x}) =  \frac{{ \sqrt{P_{\text{max}}}}}{ \eta_{\text{max}}}  \nabla \psi_{\mu}(\| \mb{x}_l \|_2 ).
\end{equation}
% \vida{the condition in the equation above should be on $P^{\text{non-linear}}_l(x)$ right? (e.g. $P^{\text{non-linear}}_l(x)>\mu$, and this translates to $\| \mb{x}_l \|^2>\mu$?) as $P^{\text{non-linear}}_l(x)$ is replacing x in (18). Also, $ \psi_{\mu}(P^{\text{non-linear}}_l (\mb{x})) =\begin{cases}
%           \frac{{ P_{\text{max}}}}{ 2\eta_{\text{max}}^2} \mb{x}_l / \mu & \| \mb{x}_l \|^2 \leq   \mu \\ 
%           \frac{\sqrt{ P_{\text{max}}\mb{x}_l}}{ \eta_{\text{max}}}  -\mu/2? &  \| \mb{x}_l \|^2 > \mu
%     \end{cases}$. If you take gradient of this, at least the first case is not as (19), for the first case the gradient would be $\frac{{ P_{\text{max}}}}{ 2\eta_{\text{max}}^2\mu}$, or am I missing something?
%     Also, a general thing about the smoothing function in (18), as I see it, when you give an argument to it, like x, at it guarantees that at $x=\mu$ it has derivative equal to 1 (right and left derivative). Your power function had gradient equal to infinity when x--->0, where did it need to get smoothed using smoothing function in(18)? I mean at which point?} 
The gradient of the penalty term does not require such smoothing and is found as follows: 
\begin{equation}
   \sum_{k=1}^K \lambda(i)  \nabla  \Psi_k(\mb{x}) =  \sum_{k=1}^K \lambda(i) [g_k(\mb{x})]_+  \nabla  g_k(\mb{x}),
\end{equation}
where the gradient of $g_k(\mb{x})$ specifically is, 
\begin{equation}
    \nabla g_k(\mb{x}) =  \left( 
            \frac{(\mb{C}_k + \mb{C}_k^T)}{2 \sqrt{\mb{x}^T \mb{C}_k \mb{x} + \sigma_{DL}^2}} 
           \mb{x}- \sqrt{\frac{1 + \gamma_k }{\gamma_k }}\tilde{\mb{b}}_k 
            \right).
\end{equation}
Finally, the full gradient is then computed as, 
\begin{equation}
    \nabla f^{\lambda(i)}(\mb{x}) = \sum_{l=1}^{L} \nabla \hat{P}^{\text{non-linear}}_l (\mb{x}) +  \sum_{k=1}^K \lambda(i) [g_k(\mb{x})]_+  \nabla  g_k(\mb{x})   .
\end{equation}
\subsection{Projection}
At every iteration, the estimate is projected back into the feasible region of $\mathcal{P}2$; this can be achieved on a per-AP basis in closed form via the solution by Bauschke \textit{et al.} \cite{Bauschke_Bui_Wang_2018}: 
\begin{equation}
\begin{aligned}    
    \mathcal{P}_{\mathcal{C}} (\mb{x}) &: \mathbb{R}^{N} \rightarrow  \mathbb{R}^{N}_+ : \\
    & \mb{x}_l \mapsto  \frac{ \sqrt{P_{\text{max}}}}{\mathrm{max}(\sqrt{P_{\text{max}}}, \|[\mb{x}_l]_+\|_2))}[\mb{x}_l]_{+} \quad \forall l.
    \end{aligned}
\end{equation}
The stepsize $\alpha_t$ is chosen via Armijo backtracking. %\cite{Armijo_1966}
This backtracking starts from a large stepsize and decreases it until a sufficient decrease in the cost function is found \cite{bentalOpt2lectures}. This sufficiency is defined by the Armijo-Goldstein inequality i.e. 
\begin{equation}
\begin{aligned}
    & f^{\lambda(i)}(\mb{x}^{t}) - f^{\lambda(i)}(\mathcal{P}_\mathcal{C}(\mb{x}^{t} - \alpha_t \nabla f^{\lambda(i)}(\mb{x}^{t}))) \\ 
    & \qquad \geq \tau \alpha_t \| \nabla f^{\lambda(i)}(\mb{x}^{t})) \|_2^2 \,.
    \end{aligned}
\end{equation}
Furthermore, we choose $\tau$ and $\zeta$ as $10^{-4}$ and $3$ respectively. 
The algorithm is terminated when the relative decrease of the objective function is smaller than the predefined threshold $\epsilon$, chosen here as $10^{-3}$. The final solution of  is denoted by $\mb{x}^{\ast}_{\text{ideal}}$ or $\mb{x}^{\ast}_{\text{non-linear}}$ when solving for the ideal (\ref{eq:paIdeal}) or the non-linear power model (\ref{eq:paReal}) respectively.

\subsection{Complexity Comparison}

Our complexity analysis is similar to the one provided in \cite{Mai_Ngo_Tran_2022} but is provided here for completeness' sake. The complexity of the SOCP-based solution is
$\mathcal{O}(\sqrt{K+L+1}K^4 L^3)$. On the other hand, the proposed method's complexity heavily depends on the number of iterations and, thus, the chosen precision. The gradient descent method's complexity depends on the gradient computation, which scales as $\mathcal{O}(L^2K^2)$ and then further scales with the number of iterations in the APG, $I_{\text{APG}}$ and penalty method, $I_{\text{penalty}}$, leading to a total complexity of $\mathcal{O}(L^2K^2 I_{\text{APG}}I_{\text{penalty}})$.

\section{Results}

To demonstrate the performance of our method, we discuss the superior scaling of our proposed method, the advantages of optimising for the actually consumed power at the PA and the induced sparsity of the solution.

% \subsection{Convergence}
% \begin{figure}[htb]
%     \centering
%     \input{figures/convergence/convergenceSEs}
%     \caption{Example of the convergence of the per-user SE to the SE constraint as a function of the iteration count in the penalty method for a total of five users, 15 access points and a target SE of 10.5 Bps/Hz. The different curves are the achieved SEs for each user at every iteration.}
%     \label{fig:constraintConvergence}
% \end{figure}
% Figure~\ref{fig:constraintConvergence} demonstrates how the users' SEs converge to the targetted SE as the penalty parameter increases and the solution converges to the solution of the constrained problem. In this work, the penalty method is stopped when each user's SE is within 1\% of the targetted SE. The result proves the penalty method's effectiveness and that $\bs{\lambda}\rightarrow \infty$ does not necessarily need to occur to approximately satisfy the constraint. In the provided example, we satisfy the constraints within 10 iterations. 

\subsection{Runtime}
\begin{figure}[htb]
    \begin{tikzpicture}
\pgfplotstableread[col sep=comma,]{figures/largeM/largeMRuntimes.csv}\datatable
\begin{axis}[
    xlabel={Number of Access Points},
    ylabel={Runtime [s]},
    legend style={
       at={(0.02,0.95)},
       anchor=north west,
       legend columns=1,
       nodes={scale=0.8, transform shape}},
    grid style={line width=.1pt, draw=gray!10},
    major grid style={line width=.2pt,draw=gray!50},
    grid style=dashed,
    grid=both,
    width=1\linewidth,
    height=0.35\linewidth,
    mark repeat=1,
    cycle list name=exotic,
    xmin=50,
    xmax=400
]    
    \addplot[mark=o, color=magenta] table [x, y={y1}]       {\datatable};
    \addlegendentry{Proposed Method}
    \addplot[mark=x, color=teal]  table [x, y={y2}]{\datatable};
    \addlegendentry{CVX}
\end{axis}
\end{tikzpicture}
    \vspace{-0.5cm}
    \caption{Comparison of runtime between the proposed method and a current state-of-the-art convex solver for a scenario with $15$ users and varying APs. \label{fig:runtime}  }
    \vspace{-0.3cm}
\end{figure}
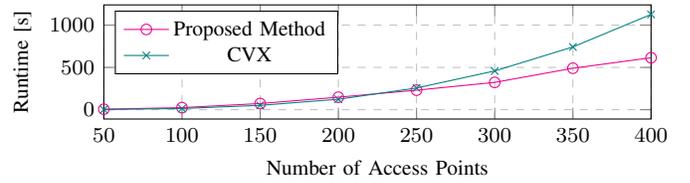
Figure~\ref{fig:runtime} highlights the significantly improved complexity of the proposed method. 
In particular, while our method is slightly slower for a network with 200 APs, it is already twice as fast for a network with 400 APs. 
Since CF mMIMO networks are expected to have many APs, this scaling is essential. 
Furthermore, since the power allocation problem should be solved quite regularly due to user mobility and changing QoS requirements, it should terminate very quickly. 
Additionally, we note that the proposed method only incurred a mean relative error of 0.21\% on the total consumed power when compared with SDPT3 via CVX.
Note that the runtime of the penalty method can be significantly improved by requiring a lower precision. 
This can be achieved by loosening the convergence criteria of the 
APG method and the penalty method.

\subsection{Power Consumption Model}
\vspace{-0.2cm}
\begin{figure}[htb]
    \centering
    \begin{tikzpicture}
  \pgfplotstableread[col sep=comma,]{figures/fractions.csv} \datatable
\begin{axis}[%
width=0.95\linewidth,
xmin=0.1,
xmax=1,
xlabel={Fraction of Max-Min Rate},
ymin=0,
ylabel={Relative Power Savings [\%]},
ymax=30,
    legend style={
       at={(0.02,0.02)},
       anchor=south west,
       legend columns=1,
       nodes={scale=0.8, transform shape}},
    grid style={line width=.1pt, draw=gray!10},
    major grid style={line width=.2pt,draw=gray!50},
    grid style=dashed,
    grid=both,
    width=1\linewidth,
    height=0.5\linewidth,
]
\addplot [ mark=o, color=magenta, mark options={solid}]
  table [x, y={y1}]{\datatable};%
\addlegendentry{Proposed Method}

\addplot [ mark=x, color=teal, mark options={solid}]
  table [x, y={y4}]{\datatable};%
\addlegendentry{CVX}

\addplot [ mark=o, color=magenta, mark options={solid}]
  table [x, y={y2}]{\datatable};%

\addplot [ mark=o, color=magenta, mark options={solid}]
  table [x, y={y3}]{\datatable};%

\addplot [ mark=x, color=teal, mark options={solid}]
  table [x, y={y5}]{\datatable};%

\addplot [ mark=x, color=teal, mark options={solid}]
  table [x, y={y6}]{\datatable};%

  \draw (0.547\linewidth,0.24\linewidth) ellipse (0.1cm and 0.2cm);
  \node[] at (0.547\linewidth,0.28\linewidth) {\footnotesize{$L = 50$}};
  
  \draw (0.547\linewidth,0.16\linewidth) ellipse (0.1cm and  0.2cm);
  \node[] at (0.547\linewidth,0.2\linewidth) {\footnotesize{$L = 25$}};
  
  \draw (0.547\linewidth,0.09\linewidth) ellipse (0.1cm and 0.2cm);
  \node[] at (0.547\linewidth,0.13\linewidth){\footnotesize{$L = 10$}};

\end{axis}

\end{tikzpicture}%
    \caption{The relative power consumption saving ((\ref{eq:relsaving})) when accurately accounting for the non-linear power consumption model in the transmit power minimization.}
    \vspace{-0.4cm}
    \label{fig:models}
\end{figure}
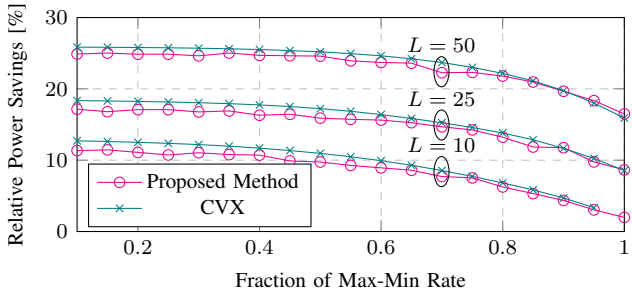

Figure~\ref{fig:models} highlights the importance of optimising for the consumed power model directly instead of just the transmitted power. The figure shows the relative consumed power savings at the PAs in the same system for different required SEs as a fraction of the max-min rate for different numbers of APs $L$. The y-axis shows the relative power savings when the non-linear power model is the true model, and the network is accurately optimised for this power consumption instead of just optimising for an ideal consumption model. The relative power saving is then computed as, 
    \begin{equation}
        % \text{Relative Power Saving} = 
        \frac{
        P^{\text{non-linear}}_{\text{total}} ( \mb{x}^{\ast}_{\text{ideal}}) -  P^{\text{non-linear}}_{\text{total}} ( \mb{x}^{\ast}_{\text{non-linear}}) }
        {
    P^{\text{non-linear}}_{\text{total}} ( \mb{x}^{\ast}_{\text{ideal}})
        },
        \label{eq:relsaving}
    \end{equation}
where $ P^{\text{non-linear}}_{\text{total}} ( \mb{x}^{\ast})$ is the total network power consumption for that solution, i.e., $P^{\text{non-linear}}_{\text{total}} ( \mb{x}^{\ast}) = \sum_{l=1}^L  P^{\text{non-linear}}_{l} ( \mb{x}_l^{\ast}) $.
For large networks (50 APs), we observe a relative power saving of 24.9\% when the targetted rate is only 10\% of the max-min rate, 
decreasing to 24.6\% for a QoS requirement of 50\% of the max-min rate. 
The saved power decreases significantly as the required capacity approaches the max-min rate. 
Interestingly, our method saves relatively more power when the network is under low demand, 
a heavily understudied operating regime for next-generation networks, due to the concavity in the power consumption curve of a real PA.
Furthermore, our proposed method only had a mean absolute error of 1.3\% when checked against the solver. 

\subsection{Sparsity}
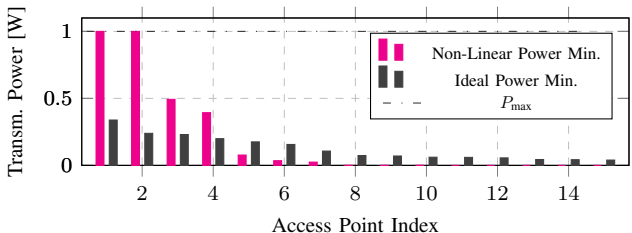
\begin{figure}[htb]
    \centering
    \begin{tikzpicture}
\pgfplotstableread[col sep=comma,]{figures/convergence/convergenceSparse.csv}\datatable
\begin{axis}[ybar,
    xlabel={Access Point Index},
    ylabel={Transm. Power [W]},
    legend style={
       at={(0.97,0.95)},
       anchor=north east,
       legend columns=1,
       nodes={scale=0.8, transform shape}},
    grid style={line width=.1pt, draw=gray!10},
    major grid style={line width=.2pt,draw=gray!50},
    grid style=dashed,
    grid=both,
    width=1\linewidth,
            bar width=3pt,
    height=0.4\linewidth,
    mark repeat=1,
    cycle list name=exotic,
             enlarge x limits=0.05,
    ymin=0,
    xmin=1,
]  
   \addplot[ybar, color=magenta, ybar legend , fill=magenta] table [x, y={y1}]{\datatable};\label{A}
    \addplot[ybar, color=darkgray,ybar legend, fill=darkgray]  table [x, y={y2}]{\datatable};
\label{B}
\end{axis}
\begin{axis}[
     xtick=\empty,
    legend style={
       at={(0.97,0.90)},
       anchor=north east,
    legend columns=1,
       nodes={scale=0.65, transform shape}},
    width=1\linewidth,
            bar width=3pt,
    height=0.4\linewidth,
    mark repeat=1,
    cycle list name=exotic,
    ymin=0,
    xmin=1,
]  
 \addlegendimage{/pgfplots/refstyle=A}
   \addlegendentry{Non-Linear Power Min.}
   \addlegendimage{/pgfplots/refstyle=B}
   \addlegendentry{Ideal Power Min.}
 \addplot[style=loosely dashdotted, black!80, domain=0:10, samples=2] {1};
    \addlegendentry{$P_{\text{max}}$}
\end{axis}
\end{tikzpicture}
    \caption{The total transmit power allocated to each AP for both power consumption models, sorted in decreasing order. This scenario considers a required SE of 6 bits/s/Hz, 5 users and 15 APs. }
    \label{fig:sparsity}
\end{figure}

Finally, Figure~\ref{fig:sparsity} shows the distribution of transmit power among the different APs for a single realisation. 
Interestingly, due to the targetted power consumption model, the PA power consumption model inherently induces sparsity on the per-AP allocated power.
 This happens because smaller transmit powers in the non-linear model are penalised more heavily, leading to a concentration of transmit powers on fewer APs. 
 This is a highly interesting result as many works in CF mMIMO attempt to shut down APs via highly complex methods, whereas here, 
 it is induced via the PA's consumption model directly. 
This insight could be leveraged to design adaptive AP switching strategies, further enhancing network energy efficiency.

\section{Conclusion}

In this work, we presented a low-complexity approach to minimizing total PA power consumption in a CF mMIMO network,  using a more realistic PA power model. Our method significantly reduces consumed power and scales more efficiently with the network size than the current state-of-the-art. We have found that by incorporating a realistic non-linear PA power consumption model, a significant amount of power can be saved (up to 24\% in the best scenario). Interestingly, more power is saved if the network is under a low traffic load. 
Additionally, the transmit power becomes concentrated on a small subset of APs in the network, which can inspire future work in AP on/off switching strategies.
Future works could also consider optimizing the smoothing variable $\mu$ based on the user requirements and the network configuration. 
Furthermore, they could explore appropriate levels of precision for the intermediate problems in the penalty method for finding 'good enough' solutions to the constrained power minimization.
\bibliographystyle{ieeetr} % We choose the "plain" reference style
\bibliography{references} % Entries are in the refs.bib file
\end{document}